\documentclass[fleqn,10pt]{wlscirep}
\usepackage[utf8]{inputenc}
\usepackage[T1]{fontenc}
\usepackage{subcaption}
\usepackage{booktabs}

\addtocontents{toc}{\protect\setcounter{tocdepth}{-1}}
\title{Automated Gleason Grading of Prostate Biopsies using Deep Learning}

\author[1,*]{Wouter Bulten}
\author[1]{Hans Pinckaers}
\author[2]{Hester van Boven}
\author[3]{Robert Vink}
\author[1]{Thomas de Bel}
\author[4]{Bram van Ginneken}
\author[1]{Jeroen van der Laak}
\author[3]{Christina Hulsbergen-van de Kaa}
\author[1]{Geert Litjens}
\affil[1]{Radboud University Medical Center, Radboud Institute for Health Sciences, Department of Pathology, Nijmegen, The Netherlands}
\affil[2]{The Netherlands Cancer Institute, Antoni van Leeuwenhoek Hospital (NKI‐AVL), Department of Pathology, Amsterdam, The Netherlands}
\affil[3]{Laboratory of Pathology East Netherlands (LabPON), Hengelo, The Netherlands}
\affil[4]{Radboud University Medical Center, Radboud Institute for Health Sciences, Department of Radiology \& Nuclear Medicine, Nijmegen, The Netherlands}

\affil[*]{wouter.bulten@radboudumc.nl}

\begin{abstract} % 100 words
The Gleason score is the most important prognostic marker for prostate cancer patients but suffers from significant inter-observer variability. We developed a fully automated deep learning system to grade prostate biopsies. The system was developed using 5834 biopsies from 1243 patients. A semi-automatic labeling technique was used to circumvent the need for full manual annotation by pathologists. The developed system achieved a high agreement with the reference standard. In a separate observer experiment, the deep learning system outperformed 10 out of 15 pathologists. The system has the potential to improve prostate cancer prognostics by acting as a first or second reader.
\end{abstract}
\begin{document}

\flushbottom
\maketitle

\thispagestyle{empty}

\section*{Introduction}
With 1.2 million new prostate cancer cases each year\cite{Bray18}, a high incidence-to-mortality ratio, and risk of overdiagnosis and overtreatment\cite{Schroder2012, Cooperberg2010}, there is a strong need for accurate assessment of patient prognosis. Currently, the Gleason score\cite{Epstein2010}, assigned by a pathologist after microscopic examination of cancer morphology, is the most powerful prognostic marker for prostate cancer (PCa) patients. However, it suffers from significant inter- and intra-observer variability\cite{Allsbrook2001a, Egevad2013}, limiting its usefulness for individual patients. It has been shown that specialized uropathologists have higher concordance rates\cite{Allsbrook2001b}, but such expertise is not widely available. Prostate cancer diagnostics thus could benefit from robust, reproducible Gleason grading at expert levels. Computational pathology and deep learning have already shown their potential in performing pathological diagnosis at expert levels in other tasks, such as identifying breast cancer metastases in lymph nodes and are inherently reproducible\cite{Ehte17}. In this study, we will investigate the potential of computational pathology and deep learning to perform automated Gleason grading of prostate biopsies. 
    
Treatment planning for prostate cancer is largely based on the biopsy Gleason score. After the biopsy procedure, tissue specimens are formalin-fixed and paraffin-embedded, cut into thin sections, stained with hematoxylin and eosin (H\&E), and examined under a microscope by a pathologist. The Gleason system stratifies the architectural patterns of prostate cancer into five types, from 1 (low-risk) to 5 (high-risk). The pathologist assigns a Gleason score, which in biopsies is a sum of the most common pattern and the highest secondary pattern, e.g., 3+5. As a result of revisions of the grading system, growth patterns 1 and 2 are currently not or rarely reported for biopsies\cite{Epst05}.

In the latest revision of the Gleason grading system the concept of five prognostically distinct grade groups was introduced\cite{Epstein2015, Epst16}; assigning scores 3+3 and lower to group 1, 3+4 to group 2, 4+3 to group 3, 3+5, 5+3 and 4+4 to group 4, and higher scores to group 5. Even though clinically relevant, initial research shows that this transition has not reduced the inter- and intra-observer variability of the grading system \cite{Ozkan2016, nagpal2019}.

Artificial intelligence, and in particular deep learning, have the potential to increase the quality of Gleason grading by improving consistency and offering expert level grading independently of geographic location. Deep learning has already shown promise in many medical fields\cite{Litj17} with examples in radiology\cite{Ardila2019}, ophthalmology\cite{Gulshan2016, DeFauw2018}, dermatology\cite{Esteva2017} and pathology\cite{Ehte17}. For prostate cancer, initially, previous research applied feature-engineering approaches to address Gleason grading\cite{Naik2007, Gertych2015, Nguyen2017}. Eventually, the field transitioned to applications of deep learning for detecting cancer\cite{Campanella2018, Litj16c}, and later Gleason grading of tissue micro arrays\cite{Arvaniti2018}, prostatectomies\cite{nagpal2019} and biopsies\cite{Lucas2019}, where the latter focuses solely on Gleason 3 vs Gleason >=4.

\begin{figure}[t!]
    \centering
    \includegraphics[width=\columnwidth]{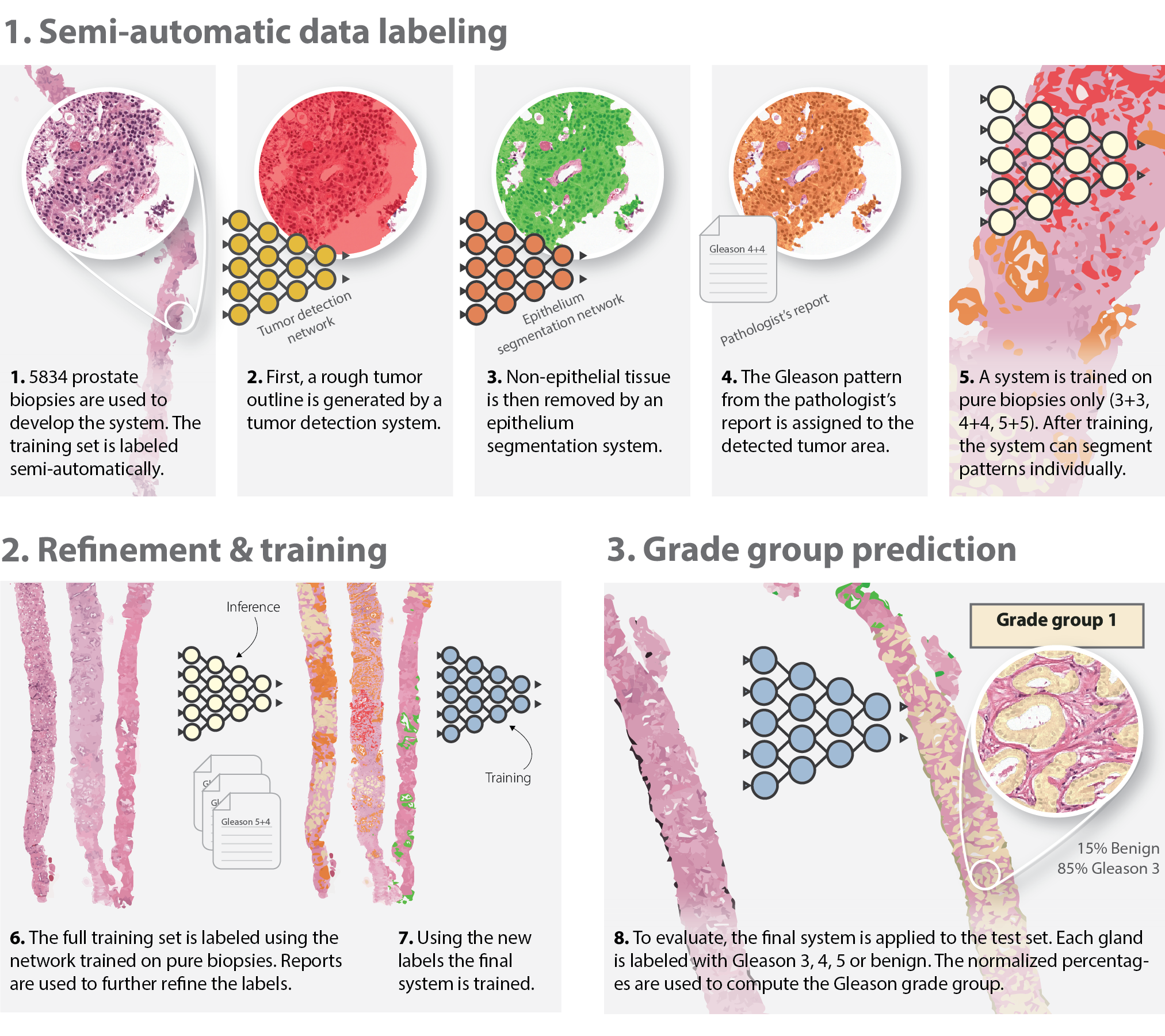}
    \caption{Overview of the development of the deep learning system. We employ a semi-automated method of labelling the training data (top row), removing the need for manual annotations by pathologists. The final system can assign Gleason growth patterns on a cell-level.}
    \label{fig:infographic}
\end{figure}

This study adds a fully-automated cancer detection and Gleason grading system for entire prostate biopsies, trained without the need for manual pixel-level annotations (Figure \ref{fig:infographic}), focusing on the full range of Gleason grades, and evaluated on a large cohort of patients with an expert consensus reference standard and an external tissue microarray test set.
 
The large inter-observer variability of Gleason grading requires a consensus reference standard. We asked three pathologists with a subspeciality in uropathology and more than 20 years of experience to grade a test set of 550 biopsies. We compared our method to the consensus grading using quadratic Cohen's kappa. The method was also compared to a larger cohort of 13 external pathologists and two pathologists in training from 10 different countries on a subset of the test set consisting of 100 biopsies. Additionally, we compared against two pathologists on a published external tissue microarray dataset previously employed to validate an automated Gleason grading system. Last, the system was validated in its ability to discriminate between benign and malignant biopsies, and identify biopsies containing aggressive cancer (grade group $>$ 1) using receiver-operating characteristic analysis.

\section*{Results}

\subsection*{Establishing the reference standard}

We collected anonymized pathology reports of 1243 patients from the pathology archives of the Radboud University Medical Center. The need for informed consent was waived by the local ethics review board (2016-2275). Of each patient we digitized a single glass slide, resulting in a total set of 1243 whole-slide images, containing 5834 biopsies. The patients were split into sets for training (933 slides, 4712 biopsies), a set for tuning (100 slides, 497 biopsies) and an independent test set (210 slides, 550 biopsies). From the test set, we selected 100 biopsies to be graded by an external panel, further referenced as the ``observer set''.

The training and tuning set combined consisted of 5209 biopsies. As manually outlining and grading all individual tumorous glands and cells in these biopsies is practically unfeasible, we employed a semi-automatic labeling technique. This technique was based on previously developed epithelium\cite{Bult19} and prostate cancer segmentation\cite{Litj16c} algorithms in combination with slide-level Gleason grades taken from the original pathology report.

The 550 biopsies of the test set were excluded for model training or hyperparameter optimization. Three pathologists specialized in uropathology, and with more than 20 years of experience each, reviewed all the biopsies from the test set individually and reported the Gleason score and grade group. For technical reasons (unrepresentative or damaged tissue, out of focus image), 15 cases could not reliably be graded and were therefore discarded.

Subsequently, a consensus score and grade group were determined for the cases in the test set and used in the remainder of the analysis as the reference standard. The inter-rater agreement of the three experts in the first round was high with an average quadratic Cohen's kappa of 0.925.

\subsection*{Overview of the deep learning system}

We developed a deep learning system using the U-Net\cite{Ronn15} architecture that predicted the grade group of a biopsy in two steps. First, the whole biopsy was segmented by assigning Gleason growth patterns to tumorous glands, and benign glands are classified as benign. From this segmentation, a normalized ratio of epithelial tissue could be calculated: \%benign, \%G3, \%G4, \%G5. In the second stage, the grade group is determined based on the normalized volume percentages of each growth pattern.

We followed the original U-Net model architecture, but after experimentation deepened the network to consist of six levels, added additional skip connections within each layer block, and used up-sampling operations in the expansion path. Experimentation showed the best performance on the tuning set using sampled patches with a size of $1024\times1024$ and a pixel resolution of $0.96\, \mu m$. Adding additional training examples from hard negative areas increased the performance of the network, especially in distinguishing between benign, inflammatory, and tumorous tissue.

\subsection*{Comparison of the deep learning system and reference standard}
\begin{figure}[t]
\centering
\begin{subfigure}{.5\textwidth}
  \centering
  \includegraphics[width=1\columnwidth]{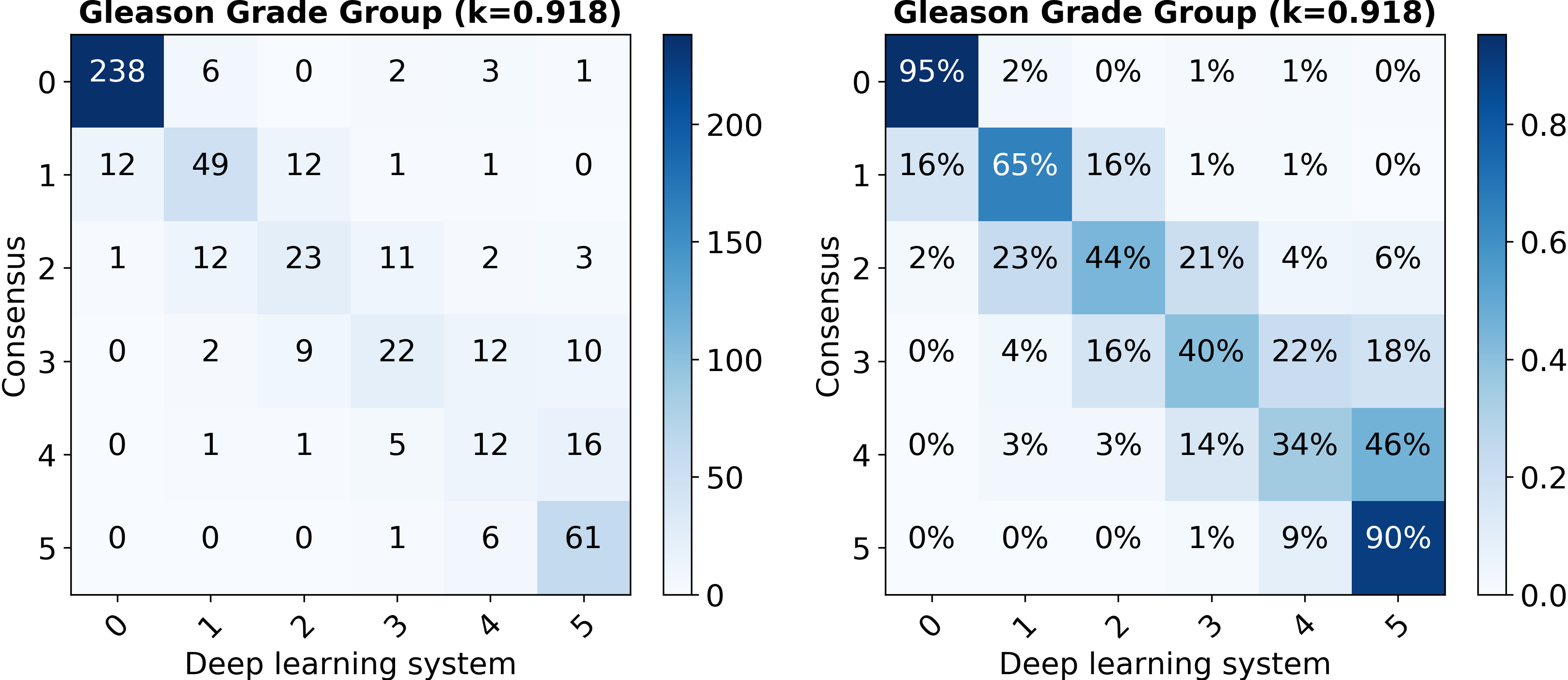}
  \caption{}
  \label{fig:confmatrix_full}
\end{subfigure}%
\begin{subfigure}{.5\textwidth}
  \centering
  \includegraphics[width=1\columnwidth]{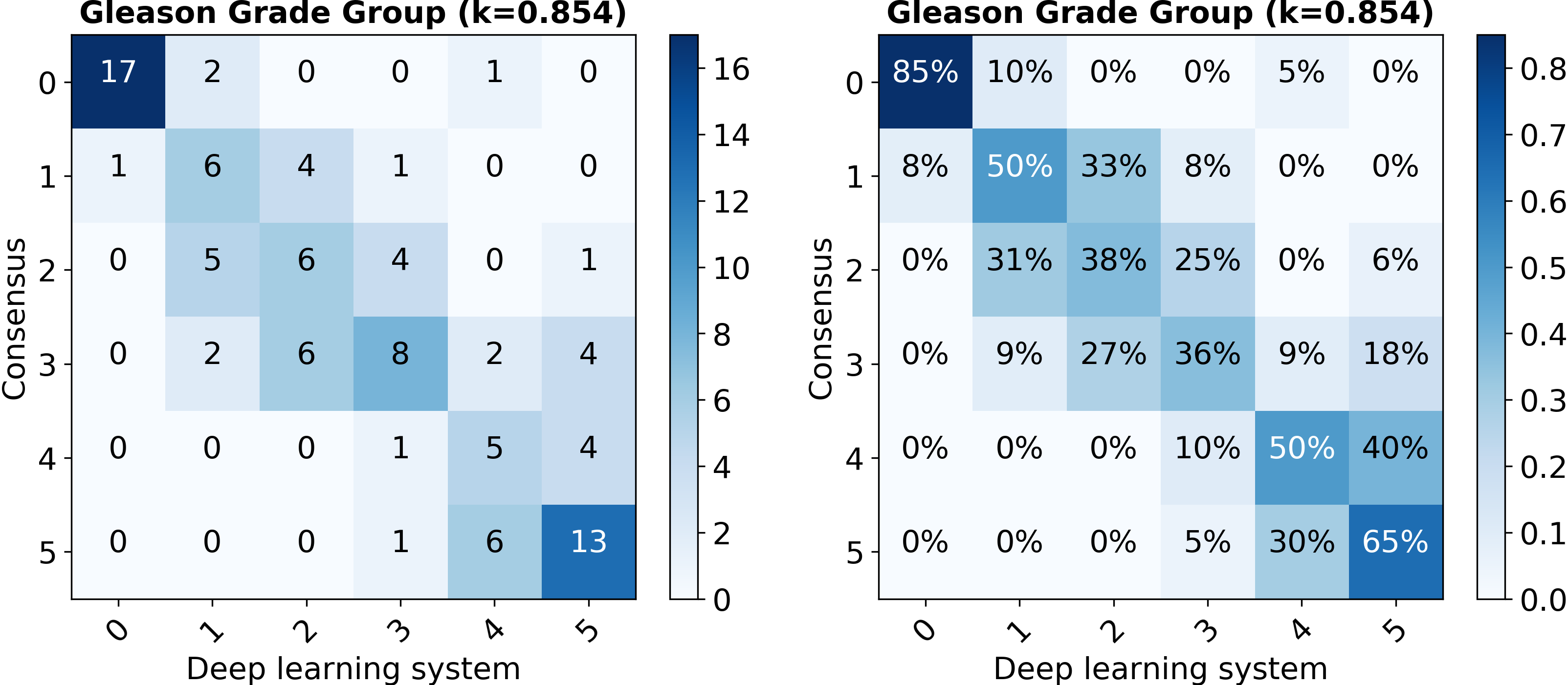}
  \caption{}
  \label{fig:confmatrix_sub}
\end{subfigure}
\caption{Confusion matrices on Gleason grade group for whole test set (a) and the observer set (b). Quadratic Cohen's kappa is shown for each set. Most errors by the deep learning system are made in distinguishing between grade group 2 and 3, and grade group 4 and 5.}
\label{fig:test}
\end{figure}

On the 535 biopsies of the test set, our deep learning system achieved an agreement of 0.918 (quadratic kappa) with the consensus grade group. Most of the discrepancies between the system's predictions and the consensus are around two decision boundaries (Figure \ref{fig:confmatrix_full}). The first group of errors occurs between grade groups 2 and 3. Visual inspection of the system outputs reveals that discrepancies result from the fact that the difference between these grade groups solely depends on the volume percentage of pattern 4. A second cut-off where the system misclassifies is between grade groups 4 and 5, often due to over grading a pattern 4 region.

\begin{figure}[t]
\centering
\begin{subfigure}{1\textwidth}
  \begin{subfigure}{.333\textwidth}
  \includegraphics[width=\columnwidth]{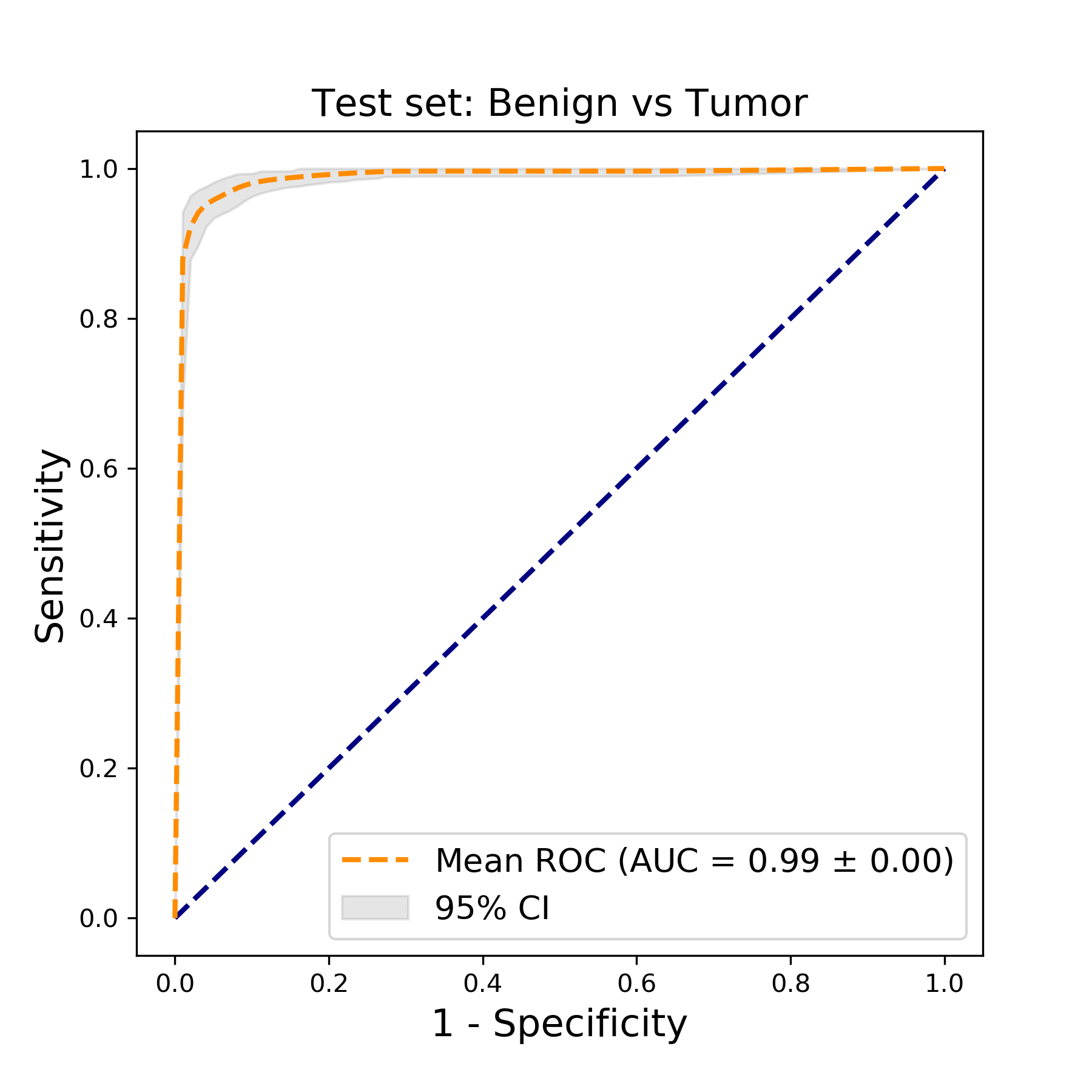}
    \caption{}
      \label{fig:roc_full_tumorbenign}
  \end{subfigure}%
  \begin{subfigure}{.333\textwidth}
    \includegraphics[width=\columnwidth]{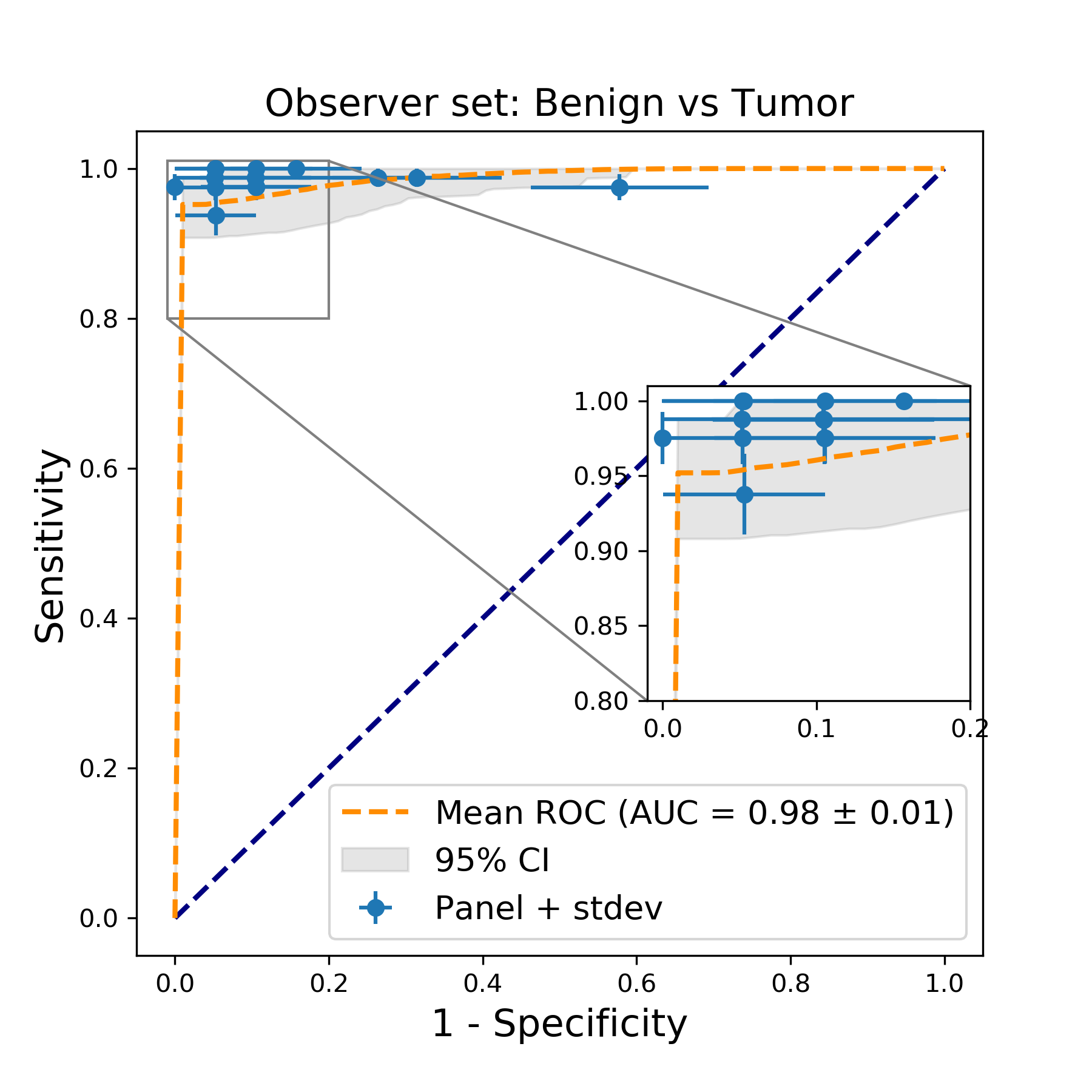}
    \caption{}
    \label{fig:roc_100_tumorbenign}
  \end{subfigure}%
  \begin{subfigure}{.333\textwidth}
    \includegraphics[width=\columnwidth]{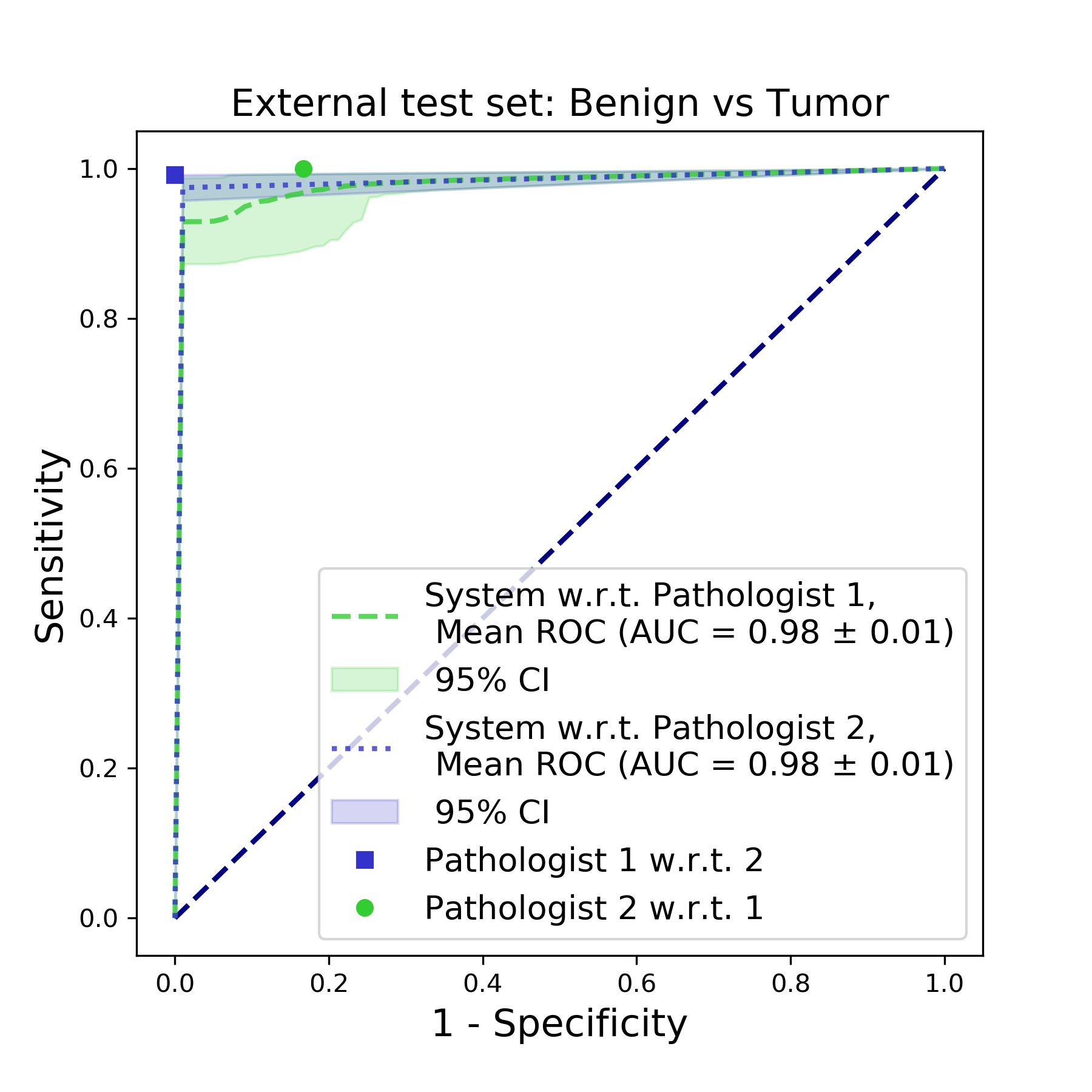}
    \caption{}
    \label{fig:roc_tma_tumorbenign}
  \end{subfigure}%
\end{subfigure}

\begin{subfigure}{1\textwidth}
  \begin{subfigure}{.333\textwidth}
    \includegraphics[width=\columnwidth]{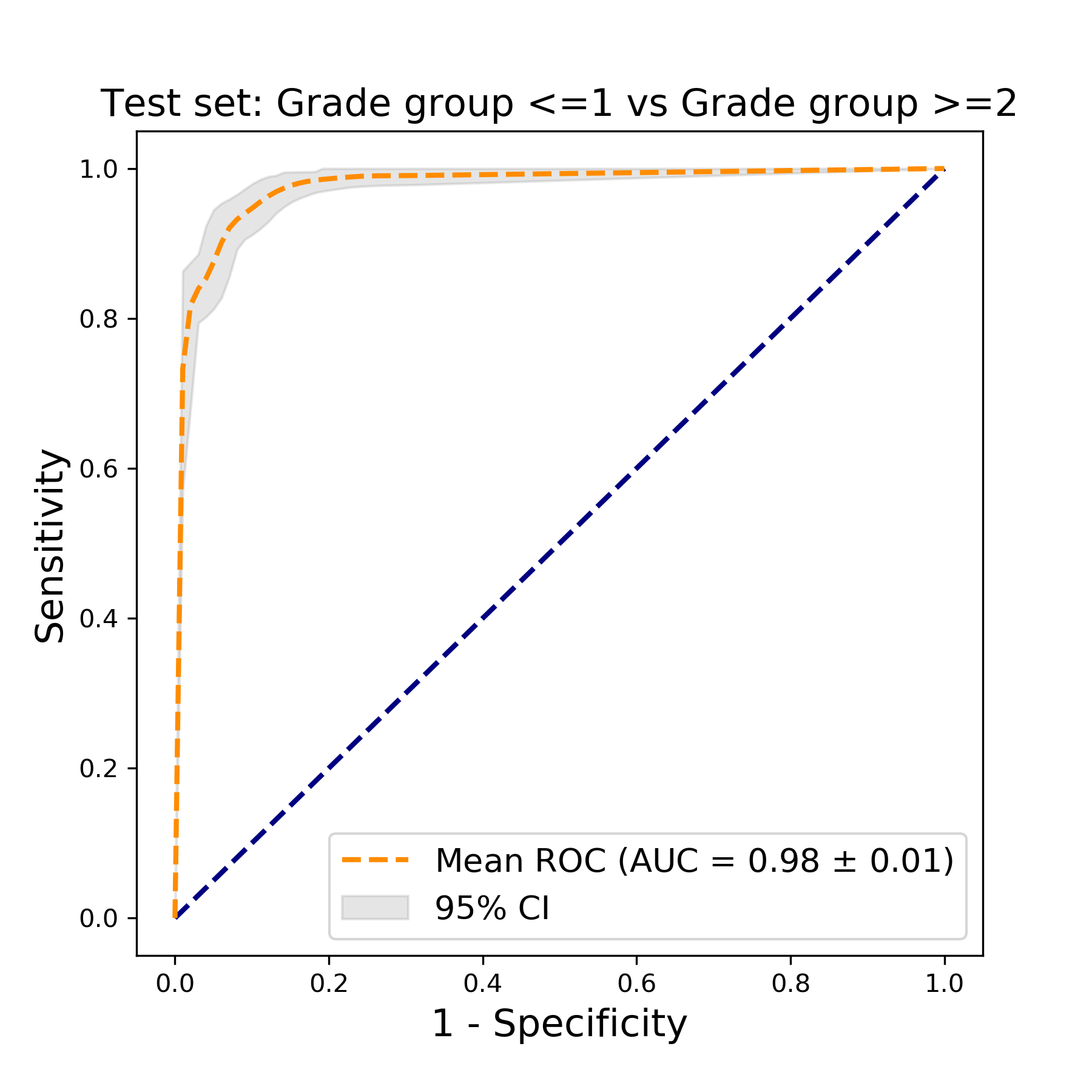}
    \caption{}
    \label{fig:roc_full_bg1vsg2}
  \end{subfigure}%
  \begin{subfigure}{.333\textwidth}
    \includegraphics[width=\columnwidth]{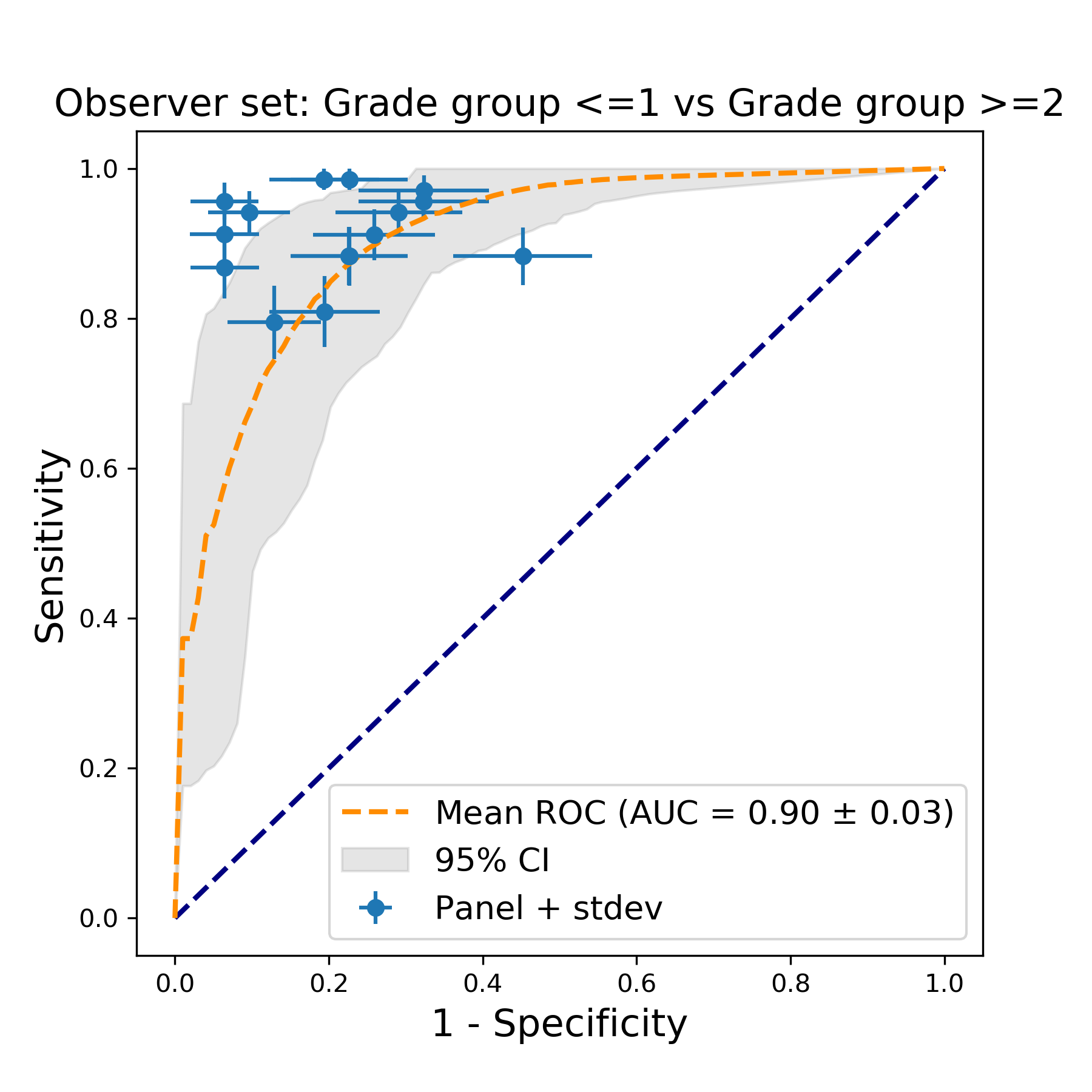}
    \caption{}
    \label{fig:roc_100_bg1vsg2}
  \end{subfigure}%
  \begin{subfigure}{.333\textwidth}
    \includegraphics[width=\columnwidth]{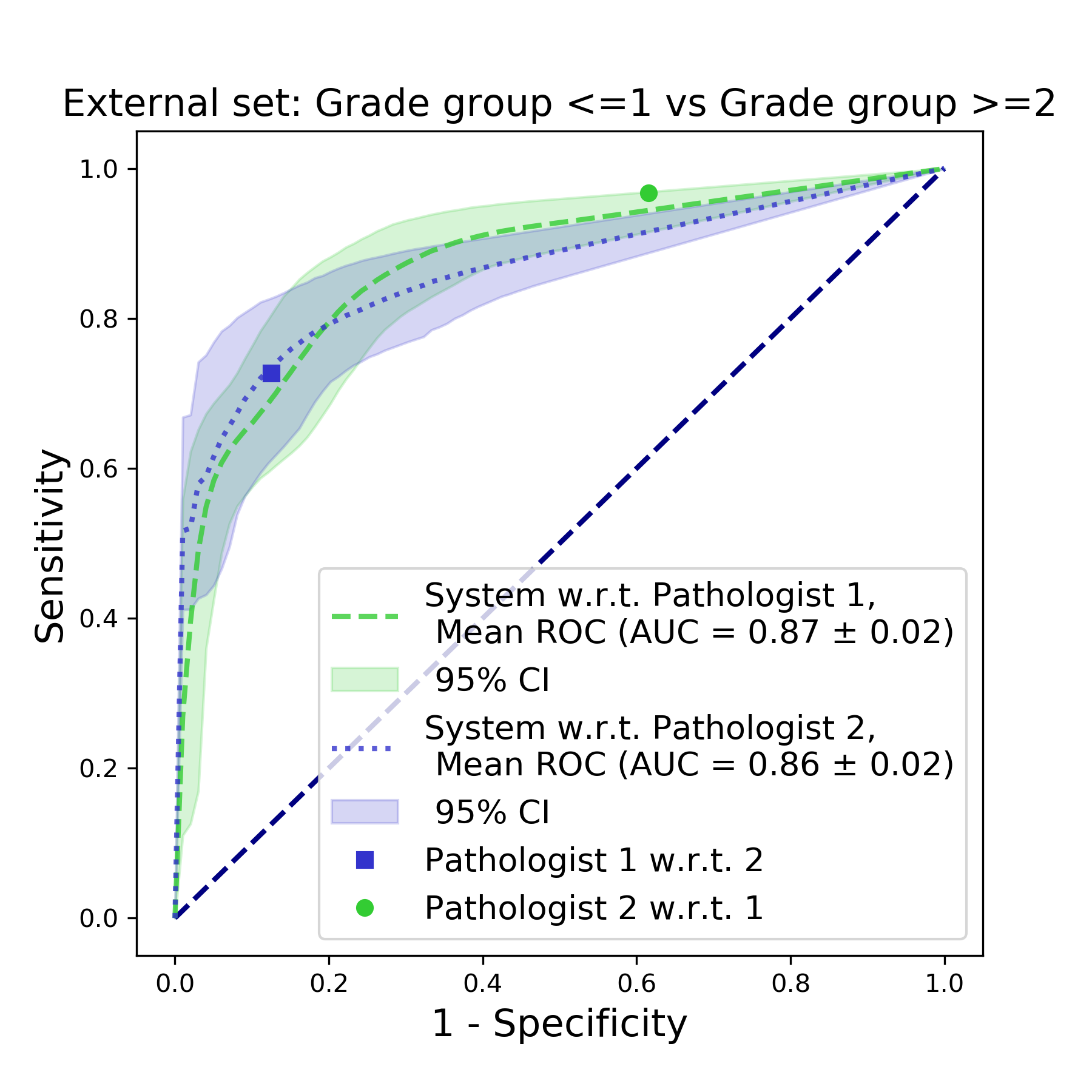}
    \caption{}
    \label{fig:roc_tma_g1vsg2}
  \end{subfigure}%
\end{subfigure}

\caption{Bootstrapped ROC analysis on two clinical relevant cutoffs: tumor versus benign (top row), and grade group $\leq 1$ versus grade group $\geq 2$ (bottom row). The first column shows results on the complete test set (535 cases). The middle column shows the results on the observer set (100 cases); the values for each panel member have been added to the graphs. The last column shows the results on the external test set, comparing the predictions of the deep learning system to the two pathologists that set the reference standard.}
\label{fig:roc}
\end{figure}

We also investigated the ability of the system to group cases in clinically relevant categories using receiver operating characteristic (ROC) analysis: benign versus malignant biopsies (Figure \ref{fig:roc_full_tumorbenign}), and using grade group 2 as a cut-off (Figure \ref{fig:roc_full_bg1vsg2}). The deep learning system obtains a high area under the ROC curve of 0.990 in determining the malignancy of a biopsy (AUC of 0.990). The system can be tuned to correctly 99\% of biopsies containing tumor while having a specificity of 82\%. The same holds for the cut-off at grade group 2 with the system achieving an AUC of 0.978 on the test set.

\subsection*{Agreement of external panel and comparison with deep learning system}

A subset of 100 cases was selected from the test set, the ``observer set'', to be assessed by an external panel of 15 observers to estimate inter-observer variability. In total, 13 pathologists and two pathologists in training from 14 independent labs and ten countries were asked to grade these biopsies independently through an online viewer. All panel members had experience with Gleason grading from clinical practice. No time restriction was given for the grading process, though we asked all members to complete the grading in six weeks.

The panel showed a median inter-rater agreement of 0.819 (quadratic kappa) on Gleason grade group with the consensus (min 0.473, max 0.947, SD 0.117). The individual scores of each panel member are shown in Figure \ref{fig:kappa_vs_panel}. The agreement on Gleason score can be found in Supplementary Figure 5.

\begin{figure}[th]
    \centering
    \includegraphics[width=1\columnwidth]{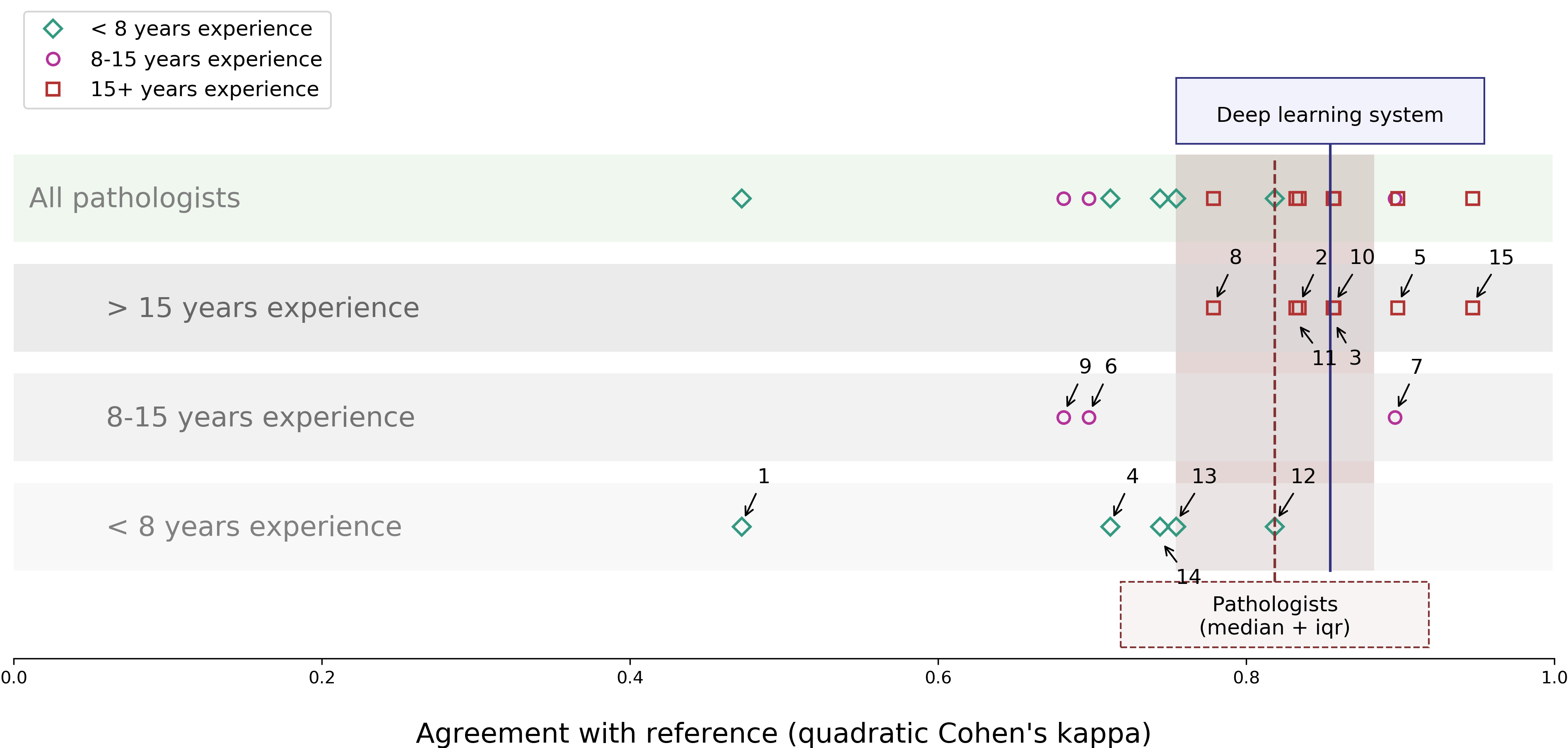}
    \caption{Agreement on Gleason grade group between each pathologist of the panel and the deep learning system with the consensus. The panel members are split out according to their experience level. Additionally, the median kappa of the pathologists is shown in brown.}
    \label{fig:kappa_vs_panel}
\end{figure}

On the observer set, the system achieves a kappa value of 0.854, which is higher than the median value of the panel. The system outperforms 10 out of the 15 panel members (Figure \ref{fig:kappa_vs_panel}). Only three panel members clearly score higher on agreement with the reference standard. The performance of the deep learning system is better than that of pathologists with less than 15 years of experience (two-sided permutation test, $p=0.036$) and scores not significantly different than pathologists with more than 15 years experience (two-sided permutation test, $p=0.955$).

The deep learning system scores better than three pathologists of the panel but lower than most on accuracy  (Supplementary Figure 6 and 7). The lower accuracy is mostly caused by one-off errors between grade groups 2 versus 3, and 4 versus 5. A two-sided permutation test on the difference between system accuracy and the median of the panel showed no significant difference ($p = 0.149$). A few cases of the observer set, including predictions by the system, are displayed in Figure \ref{fig:overlay}. Additional examples of the system output are displayed in Supplemental Figure 8.

\begin{figure}[h!]
    \centering
    \includegraphics[width=1\textwidth]{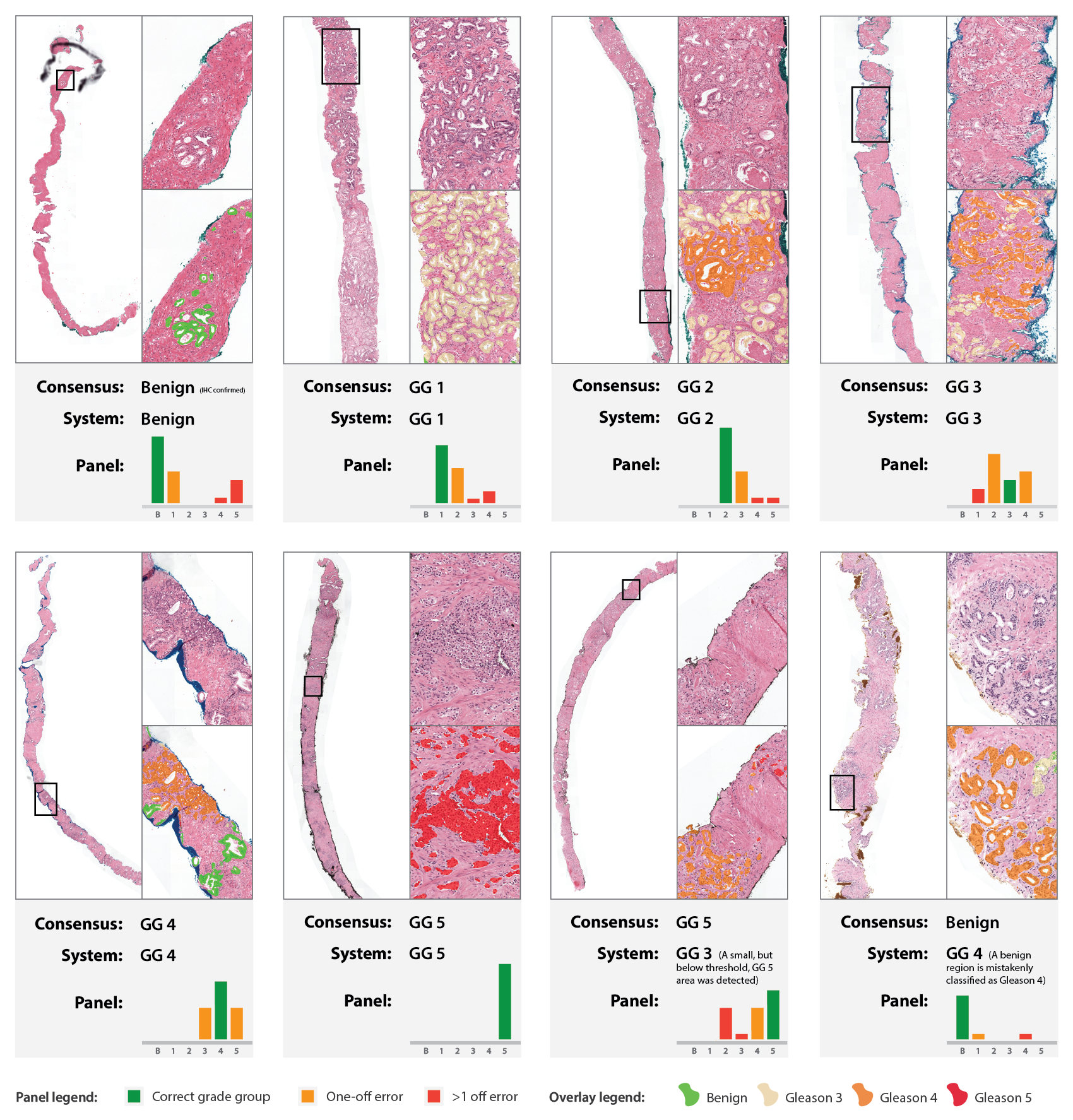}
    \caption{Examples from the observer set. For each case, the grade group of the reference standard, the predicted grade group by the deep learning system, and the distribution of grade groups from the panel is shown. The results of the panel are color-coded: green for agreement with the reference, orange for at most one deviation and red for larger errors. In the zoomed regions the Gleason prediction of the system is shown as an overlay on the tissue. The system labels each epithelial gland with either benign (green), Gleason 3 (yellow), Gleason 4 (orange) or Gleason 5 (red). The first six cases show examples of each grade group. The last two cases (bottom row) show examples of failure cases where the system predicted the wrong grade group for the case.}
    \label{fig:overlay}
\end{figure}

\begin{figure}[th]
    \centering
    \includegraphics[width=\textwidth]{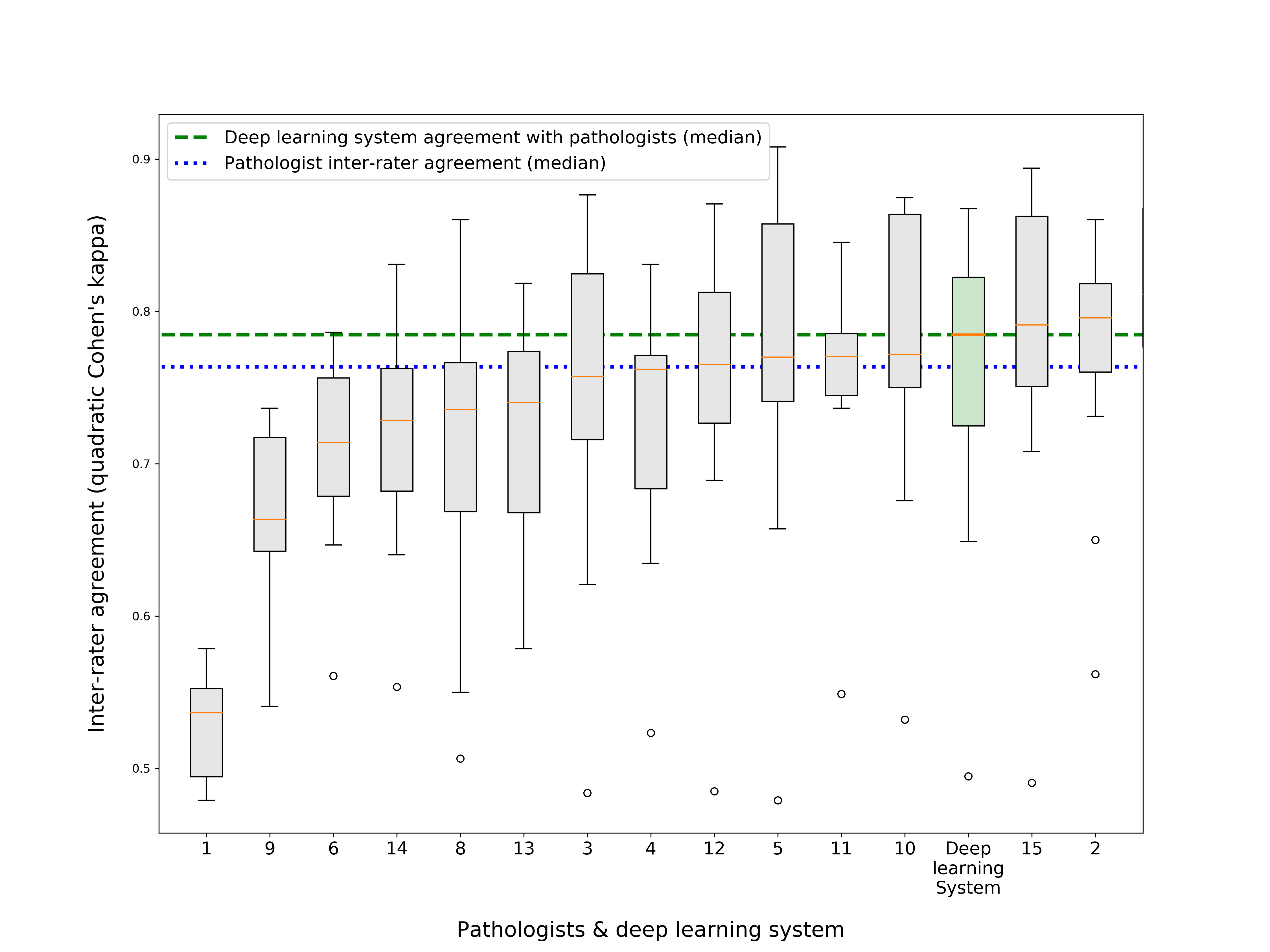}
    \caption{Inter-rater agreement between panel members. For each pathologist, the inter-rater agreement with each other pathologist from the panel was calculated. Additionally displayed is the agreement of the deep learning system with the pathologists from the panel. The pathologists and deep learning system are ordered based on their respective median agreement values. The two horizontal lines display the median agreement of all pathologists (in blue) and median agreement of the system (in green).}
    \label{fig:kappa_internal_panel}
\end{figure}

To exclude any bias in our results towards the three consensus experts, we also computed the inter-rater agreement between all pathologists independently of the reference standard. We then computed the agreement of the system with all members of the panel. Sorted by the median kappa value, the deep learning system ends in the top-3 (Figure \ref{fig:kappa_internal_panel}, Supplementary Figure 9).

As for the full test set, we also computed the ability of both the system and the panel to group cases in clinical relevant categories via ROC analysis. The deep learning system performs comparable to the panel on both tumour versus benign (AUC 0.984, Figure \ref{fig:roc_100_tumorbenign}), and grade group 2 or higher (AUC 0.904, Figure \ref{fig:roc_100_bg1vsg2}). A two-sided permutation test on the median F1 score showed no statistically significant difference between the deep learning system and the panel for both malignant versus benign ($p=0.704$) and grade group 2 as a cut-off ($p=0.838$).

\subsection*{Results on the external dataset}

We evaluated our system's generalization ability on an external dataset of tissue microarrays\cite{Arvaniti2018}. This dataset consists of two sets: 641 cores annotated by a single pathologist and used for the development of an algorithm by Arvaniti et al.,\cite{Arvaniti2018} and a second set of 245 cores used as a test set which was annotated independently by two pathologists. There was no final consensus available of the two pathologists, instead, we evaluated our method using both pathologists as the reference standard. Concerning the first pathologist, the system obtained a 0.711 quadratic kappa on Gleason score; for the second pathologist, a quadratic kappa of 0.639. The values on the test set are lower than attained by the algorithm by Arvaniti et al. (kappa of 0.71 and 0.75 for pathologist 1 and 2, respectively), but within inter-observer variability. The confusion matrices are displayed in Supplementary Figure 10.

Additionally, we tested the ability of the deep learning system to group cases into low or high risk. Using the first pathologist as the reference, the system achieves an AUC of 0.977 on distinguishing between benign and malignant biopsies (Figure \ref{fig:roc_tma_tumorbenign}) and an AUC of 0.871 using grade group 2 as a cut-off (Figure \ref{fig:roc_tma_g1vsg2}). For the second pathologist, these AUC values are similar with 0.982 and 0.861.

\section*{Discussion}

We have developed a fully automated method to grade prostate biopsies and have shown that this method can achieve pathologist-level performance on both the internal and external test sets. Compared to a consensus reference set by three expert uropathologists, the deep learning system achieved a high agreement (quadratic kappa of 0.918). Compared to an external panel of pathologists with varying levels of experience, the deep learning system outperformed 10 out of 15 panel members in quadratic kappa. Last, on the external test set, the system achieved a quadratic kappa on Gleason score of 0.711 and 0.639, which constitutes a substantial agreement, is within expected inter-observer variability, and comparable to the inter-observer agreement of the reference standard (kappa of 0.71). 

The precise gland-level segmentations of the deep learning system made it possible to compute the volume percentages of growth patterns, and we used these volumes to determine the grade group directly. We chose deliberately for an interpretable and straightforward method, instead of using an additional learned model on top of the deep learning system. Our method is similar to the method used by pathologists in clinical practice, significantly increasing the interpretability of our system.

Our method was trained on data which was labeled automatically in a semi-supervised way. By using an automated method of labeling the training data, we saved resources that would otherwise have been spent in manually labeling. Moreover, it is often practically unfeasible to precisely annotate the vast amounts of data required for deep learning, even though unannotated or sparsely annotated data is often readily available in pathology archives. The downside of the labeling method was the introduction of a high amount of label noise, shown by Cohen's kappa of 0.853 between the training labels and the reference standard. The system showed the ability to overcome this noise in the training data, given a kappa of  0.918.

Given the high incidence of prostate cancer, reducing workload for pathologists is of great clinical value. On the test set our deep learning system achieves an AUC of 0.990 on determining the malignancy of a biopsy, on the observer set an AUC of 0.984 Furthermore, the system can be tuned to achieve a sensitivity of 99\%. As such, our system could be implemented as a pre-screening tool within pathology labs, giving priority to high-grade biopsies. Besides a biopsy-level grade group, the deep learning system also outputs gland-level Gleason growth pattern overlays. These overlays can be used by pathologists as a second read, to confirm that no tumor was missed and to assist in accurately estimating growth pattern volumes. 

For prostate cancer patients, the tumor grade of a biopsy is one of the most essential parts of treatment planning. Often not only the presence but the aggressiveness of the tumor determines which treatment is advised. We showed that the deep learning system is able to group cases into relevant categories. On the test set, the deep learning system reaches an AUC of 0.978 when using grade group 2 as a cut-off. On the observer set, the deep learning system achieves an AUC of 0.904, comparable to the performance of pathologists. 

Our results extend previous work on prostate cancer detection and automated Gleason grading. Litjens et al.\cite{Litj16c} and Campanella et al.\cite{Campanella2018} showed the use of deep learning for detecting tumor in prostate biopsies. Arvanti et al. and Nagpal et al. showed the effectiveness of using deep learning for automated Gleason grading on microarrays and prostatectomies, respectively. Lucas et al.\cite{Lucas2019} showed the use of deep learning for Gleason grading on biopsies, but were limited by a small set and did not include all Gleason grades. We extend on this work by focusing on automated Gleason grading for prostate biopsies, the most important histological analysis for prostate cancer patients. Furthermore, by including both benign biopsies and biopsies from the full spectrum of Gleason grades, we created a system that is usable as a pre-screening tool and as a second-reader.

Given the reported inter- and intra-rater agreement of Gleason grading\cite{Allsbrook2001a, Allsbrook2001b, Egevad2013}, the performance of the deep learning system could only reliably be assessed through the use of an expert reference standard. We asked three expert uropathologists to grade the complete test set, which resulted in a minimum of three independent reads for every case. To further strengthen the analysis, we invited a panel of independent pathologists and pathologists in training to grade our observer set. Using this external panel we were able to show the pathologist-level performance of our deep learning system.

Before our system can be used in clinical practice, some limitations have to be addressed. The data that was used to develop the deep learning system originated from a single center. Although data augmentation was used, the performance of the system is lower on the external test set. While the performance is within the range of inter-observer variability on the internal test set, including data from multiple centers, with different staining protocols and whole-slide scanners, could further increase the robustness of the system.

Second, in this study, we focus solely on the grading of acinar adenocarcinoma in prostate biopsies, though other tumor types and foreign tissue can be present in prostate biopsies, e.g., colon glands which should be identified and excluded for grading. Additionally, other prognostic information could be present in the biopsies that we currently do not extract. As an example, we did not explicitly focus on the detection of intraductal carcinoma, which has been shown to be prognostically  relevant\cite{Kweldam2016, Kato2016}.

\section*{Methods}

\subsection*{Data acquisition}

The digital patient records of the Radboud University Medical Center were queried, and all pathologist reports for patients that underwent a prostate biopsy between 2012 and 2017 were retrieved (IRB number 2016-2275, Supplementary Figure 1). The reports were anonymized, and a text-search was employed to determine the highest mentioned Gleason score in each report. Patient reports were then sampled uniformly, stratifying by the Gleason score. Each pathology report was manually read, and for each patient, a single H\&E stained glass slide containing the most aggressive or prevalent part of the tumor was selected for scanning. Additional cases were selected containing only benign biopsies. Patients that underwent (neo-)adjuvant therapy were excluded.

The selected glass slides were scanned using a \textit{3DHistech Pannoramic Flash II 250 scanner} at 20x magnification (pixel resolution $0.24\, \mu m$) resulting in 1410 whole-slide images (WSI). Each WSI contained one to six unique biopsies, commonly with two sections per biopsy. After scanning, trained non-experts evaluated all glass slides and coarsely outlined each biopsy. Furthermore, each biopsy was assigned a Gleason score or the label `negative' based on the pathology report. In total 5834, individual biopsies were outlined, originating from 1243 WSIs. 167 scans were excluded due to several reasons, including: the scanned slide was out of focus; biopsies were not completely scanned or missed by the scanner; neo-adjuvant therapy was missed in the first assessment of the report; the Gleason score could not be assigned due to an inconclusive report; there was a mismatch between the scanned biopsies and the description of the report.

\subsection*{Collection of the reference standard for the independent test set}
From the dataset, we randomly selected 210 patients to form our independent test set (or holdout set); for each patient we included one whole-slide image. The patients included in the test set were independent of the patients in the training set and not used for the development of our system. We selected patients based on Gleason grade group, resulting in 59 glass slides containing benign biopsies and 151 malignant. From these glass slides, 550 unique biopsies were individually extracted, shuffled, and assigned a new identifier. If multiple sections of a biopsy were present on a slide, only one of those was included. To create a strong reference standard, we asked three pathologists with a sub-specialty in uropathology (C.H.v.d.K, H.v.B, R.V.) to grade the 550 biopsies individually through an online viewer, PMA.view (Pathomation, Berchem, Belgium), using the ISUP 2014 guidelines\cite{Epstein2015}.

The reference standard for the independent test set was determined in three rounds. Consensus was determined based on the reported Gleason scores and grade groups reported by the three pathologists. In the first round, each pathologist reviewed the cases individually. In the case of a positive biopsy each pathologist was asked to report: primary Gleason grade, secondary Gleason grade, tertiary Gleason grade (if present), an estimate of the total tumor volume, an estimate of the tumor volumes for the growth pattern and the Gleason grade group. Biopsies that could not be graded, e.g. due to artifacts, were flagged (Supplementary Table 1).

After the first round, 333 cases (61\%) had a complete consensus on both the grade group and Gleason score. For cases with only a slight difference between the three pathologists, the majority vote was taken (139 cases, 25\%). This was used for cases with agreement on grade group but a difference in Gleason pattern order (e.g., 5+4 versus 4+5), or a max deviation of 1 in grade group by a single pathologist. Cases with a disagreement between a positive and benign label were always flagged. For some cases, immunohistochemistry was available in the original report and, if present, was used to determine the benign/malignant label. Fifteen cases (3\%) were excluded because they could not be reliably graded.

In the second round, each biopsy without consensus in the first round (63 cases, 11\%) was presented to the pathologist whose score differed from the majority and re-graded. Additional to the pathologist's initial examination, the Gleason scores of the other pathologists were appended anonymously. Based on a re-evaluation of the biopsy and the other scores, each pathologist could update the initial examination. 

Biopsies without consensus after round two (27 cases, 5\%) were presented to the three experts in a consensus meeting (Supplementary Table 2). In the consensus meeting, the biopsies were discussed, and a final consensus score was determined. The consensus grade group distribution of the test set is displayed in Supplementary Figure 2. The confusion matrices of the first read of the three experts compared to the final reference standard are shown in Supplementary Figure 3 and 4.

\subsection*{Collection of the observer set}

From the independent test set, 100 biopsies were selected to be presented to a larger group of pathologists in an observer experiment. One of our expert pathologists (C.H.v.d.K.) selected 20 benign cases manually, controlling for a broad range of tissue patterns including, e.g., inflammation and (partial) atrophy. The remaining 80 biopsies were randomly selected, stratified for Gleason grade group based on the reported values of the same pathologist.

The 100 biopsies were made available through an online viewer and distributed to a panel of external observers. In total, 13 pathologists and two pathologists in training, from 14 independent labs and ten countries individually graded all 100 biopsies following the ISUP 2014 guidelines\cite{Epstein2015}. All members of the panel had experience with Gleason grading, but with a varying amount of years of experience. In the case of a positive biopsy each pathologist was asked to report: primary Gleason grade, secondary Gleason grade, tertiary Gleason grade (if present), an estimate of the total tumor volume, an estimate of the tumor volumes for the growth patterns and the Gleason grade group. 

\subsection*{Collection of the reference standard for the training set}

Slides from patients that were not included in the independent test set were used for training. Of these slides, 933 were used for training and 100 as a tuning set for hyperparameter and model selection. The slides from the training set contained 4712 biopsies, the tuning set 497. 

The data acquisition procedure resulted in outlined biopsies with a single label per biopsy. This labeling technique was not perfect; in some cases, inconsistent or concise pathologist reports made it difficult to match the Gleason score to a particular biopsy.

The quality of the labeling of the training set was determined by labeling the cases in the test set using the same automated method. We compared the retrieved Gleason scores of the test set with the final consensus score of the experts. The kappa values of this comparison acted as a measure of label quality. On Gleason score, the accuracy of the retrieved labels versus the reference is 0.675 (quadratic Cohen's kappa 0.819), on grade group 0.720 (quadratic Cohen's kappa 0.853); which indicates that although the agreement is reasonable, label noise is introduced in the training set. However, it is well known that deep learning systems can handle significant amounts of label noise\cite{Rolnick2017}.

 To train the deep learning system to segment individual glands, more detailed annotations than slide-level labels were required. We preprocessed the biopsies in four steps. First, tissue was automatically distinguished from background using a tissue segmentation network\cite{Band17}. Within tissue areas, a trained tumor detection system\cite{Litj16c} was applied to define a rough outline of the tumor region for each biopsy. The outlined tumor regions still contained large areas of stroma, inflammation, or other non-epithelial tissue as it was trained with coarse annotations. To refine the tumor masks, each biopsy was processed by an epithelial tissue detection system\cite{Bult19}, and after which tissue that was detected as non-epithelial tissue was removed from the tumor mask. All detected tumor tissue was assigned a label based on the Gleason score retrieved from the pathology report.
 
 We first trained a deep learning system only on biopsies with a pure Gleason score (3+3, 4+4, 5+5)\cite{Bult19a}. After training, this initial system was applied to the complete training set. This procedure made it possible to semi-automatically annotate biopsies with mixed Gleason growth patterns in the training set. Using the pathologist reports, the output was further automatically refined by removing clearly incorrect label assignments. Furthermore, any tissue originating from benign biopsies detected as malignant was relabelled as hard-negative to be over-sampled during training. A connected components algorithm was applied to make sure that each gland was assigned to a single class: benign, Gleason 3, Gleason 4, Gleason 5, or hard-negative.
 
\subsection*{Architecture of the deep learning system}
 
Our deep learning system consisted of a U-Net\cite{Ronn15} that was trained on patches extracted from the full training set. After the automatic labeling process, the system could be trained on all biopsies, including those with mixed Gleason growth patterns. Additional patches were sampled from the hard-negative areas to improve the system's ability to distinguish tumor from benign tissue. 

We followed the original U-Net model architecture but experimented with different configurations. Given the importance of morphological features in Gleason grading, we focused on multiple depths of the U-Net model, combined with different pixel spacings for the training data. For all tested networks, we added additional skip connections within each layer block and used up-sampling operations in the expansion path. Adam optimization was used with $\beta_1$ and $\beta_2$ set to $0.99$, a learning rate of $0.0005$ and a batch size of 8. The learning rate was halved after every 25 consecutive epochs without improvement on the tuning set. We stopped training after 75 epochs without improvement on the tuning set.

The network was developed using Keras\cite{chollet2015keras} and TensorFlow\cite{tensorflow2015}. Data augmentation was used to increase the robustness of the network. The following augmentation procedures were used: flipping, rotating, scaling, color alterations (hue, saturation, brightness, and contrast), alterations in the H\&E color space, additive noise, and Gaussian blurring.

\subsection*{Evaluation of the deep learning system}

After training, the deep learning system was applied to all biopsies from the test set. This procedure resulted in a label for each pixel of the slide. The total surface area of benign and tumorous tissue (all pixels assigned grade $>=$ 3) was calculated by counting the number of pixels assigned to each group. The values were normalized relative to the total volume of the epithelium, resulting in four volume percentages for each biopsy: \%benign, \%G3, \%G4, and \%G5.

The volume percentages were used to assign a Gleason score and Gleason Grade group based on the guidelines for biopsies in clinical practice\cite{Epstein2015}. Based on the tuning set, we determined the optimal decision thresholds for the deep learning system. Biopsies with less than 10\% of the epithelium predicted as tumor tissue were considered benign. For positive biopsies, the Gleason score was assigned based on the primary and secondary pattern given the latter occupied at least 7\% of the epithelium, otherwise, only the primary pattern was used (e.g., 3+3 or 4+4). In the presence of a higher tertiary pattern, the tertiary pattern was used instead of the secondary pattern. From the Gleason score, the grade group was determined. Further details regarding the assignment of grade groups can be found in the supplemental material.

\subsection*{Evaluation on the external dataset}

We evaluated the system on an external dataset of tissue micro-arrays\cite{Arvaniti2018} to assess the robustness of the system to data from a different center. It consists of two sets of tissue cores: 641 cores annotated by a single pathologist and used for training and algorithm development in the original paper. The second set of 245 cores was used for testing and annotated and graded by two pathologists (inter-rater agreement quadratic Cohen's kappa of 0.71).

To account for stain and scanner variations, we applied an unsupervised normalization algorithm based on CycleGANs in combination with Gaussian blurring\cite{bel19a}. A brief description of the algorithm can be found in the supplemental material. After normalization of the external test images, our deep learning system, without any modification, was applied to the 245 normalized test images.

 The original paper reports the quadratic kappa on Gleason score as the primary metric. The Gleason scores were determined using the standard for grading prostatectomies: the sum of the most and second most common growth patterns. We applied our algorithm to the test set of this paper and computed the quadratic kappa on Gleason score to allow a one-to-one comparison of our algorithm to the algorithm developed by Arvaniti et al. To account for the difference in the grading systems between prostatectomies and biopsies we adjusted the decision thresholds of the deep learning system using the training data set of 641 cores, resulting in a tumor threshold of 1\% and a secondary pattern threshold of 2\%.
 
\subsection*{Statistical analysis}
 
To compare the performance of the deep with the external panel of pathologists, we performed multiple permutation tests. The permutation test was implemented as follows: For each case in the observer set, we obtained a list of predictions, one by the deep learning system and the remainder by the pathologists. In each iteration of the permutation test, for each case, we swapped the grade group prediction of the system with a random prediction from the list of predictions. After swapping the predictions, the test statistic was computed. Repeating this procedure 10.000 times resulted in a null distribution of the test statistic. The original test statistic was then compared to the null distribution, resulting in a two-tailed p-value. 

For the analysis on agreement with the reference, we defined the test statistic as the difference between the kappa of the deep learning system and the medium kappa of the pathologists. The panel of pathologists was split into two groups, those with less than 15 years experience and those with more, and a permutation test was performed for both groups. The analysis of the ROC curves was done using the difference in  F1-score as the metric.  The decision threshold for the system was based on the point that maximized the AUC. The comparison of grade group accuracy was made using the difference in the accuracy of the system and the median accuracy of the pathologists.
 
\bibliography{bibliography}

\section*{Acknowledgments}

This study was financed by a grant from the Dutch Cancer Society (KWF), grant number KUN 2015-7970.
\mbox{}\\

\noindent The authors would like to thank the following pathologists and pathologists in training for participating in our study as part of the panel:

\begin{itemize}
    \item Paulo G. O. Salles, M.D. Ph.D. MBA, Instituto M\'ario Penna, Belo Horizonte, Brazil;
    \item Vincent Molini\'e, M.D., Ph.D., CHU de Martinique, Universit\'e des Antilles, Fort de France, Martinique;
    \item Jorge Billoch-Lima, M.D. FCAP, HRP Labs, San Juan, Puerto Rico;
    \item Ewout Schaafsma, M.D., Ph.D., Radboud University Medical Center, Nijmegen, The Netherlands;
    \item Anne-Marie Vos, M.D., Radboud University Medical Center, Nijmegen, The Netherlands;
    \item Xavier Farr\'e, M.D., Ph.D., Department of Health, Public Health Agency of Catalonia. Lleida, Catalonia, Spain.
    \item Awoumou Belinga Jean-Joël, M.D., Ph.D. Candidate, Department of Morphological Sciences and Anatomic Pathology, Faculty of Medicine and Biomedical Sciences, University of Yaounde 1, Cameroon;
    \item Jo\"elle Tschui, M.D., Medics Pathologie, Bern, Switzerland;
    \item Paromita Roy, M.D., Tata Medical Center, Kolkata, India;
    \item Em\'ilio Marcelo Pereira, M.D., Oncoclínicas group, Brazil;
    \item Asli Cakir, M.D., Pathologist, Istanbul Medipol University School of Medicine, Pathology Department, Istanbul, Turkey;
    \item Katerina Geronatsiou, M.D., Centre de Pathologie, Hopital Diaconat Mulhouse, France;
    \item G\"unter Saile, M.D., Histo- and Cytopathology, labor team w ag, Goldach SG, Switzerland;
    \item Am\'erico Brilhante, M.D., Salom\~aoZoppi Diagnostics, S\~ao Paulo, Brazil;
    \item Guilherme Costa Guedes Pereira, M.D., Laboratory Histo Patologia Cir\'urgica e Citologia, Jo\~ao Pessoa-PB, Brazil. 
\end{itemize}

Furthermore, the authors would like to thank Jeffrey van Hoven for assisting with the data collection and scanning, and Milly van Warenburg, Nikki Wissing and Frederike Haverkamp for their help making the manual annotations.

\section*{Author Contributions}

W.B. performed the data selection, experiments, analyzed the results and wrote the manuscript. H.P. was involved with the data collection and experiments. H.v.B., R.V. and C.H.-v.d.K. graded all cases in the test set. T.d.B. was involved with the application of the method to the external data. G.L., C.H.-v.d.K., J.v.d.L. and B.v.G. supervised the work and were involved in setting up the experimental design. All authors reviewed the manuscript and agree with its contents.

\section*{Additional information}

\textbf{Competing interests}

\noindent W.B., H.P., T.d.B., C.H.-v.d.K., R.V., H.v.B., and B.v.G declare no conflict of interest.

\noindent J.v.d.L. is member of the scientific advisory boards of Philips, the Netherlands and ContextVision, Sweden and receives research funding from Philips, the Netherlands and Sectra, Sweden.

\noindent G.L. received research funding from Philips Digital Pathology Solutions (Best, the Netherlands) and has a consultancy role for Novartis (Basel, Switzerland. He received research grants from the Dutch Cancer Society (KUN 2015-7970), from Netherlands Organization for Scientific Research (NWO) (project number 016.186.152) and from Stichting IT Projecten (project PATHOLOGIE 2)

\end{document}